\documentclass[chicago]{emulateapj}
\usepackage{amsmath}
\usepackage{graphicx}
\usepackage[dvips]{color}

\slugcomment{Draft v4 -- 14Nov2015}
\shortauthors{}

\begin{document}

\title{Absence of Significant Cool Disks in Young Stellar Objects Exhibiting Repetitive Optical Outbursts}

\author{Hauyu Baobab Liu\altaffilmark{1,2}} \author{Roberto Galv\'{a}n-Madrid\altaffilmark{3}} \author{Eduard I. Vorobyov\altaffilmark{4,5}}  \author{{\'A}gnes K{\'o}sp{\'a}l\altaffilmark{6}} \author{Luis F. Rodr\'{i}guez\altaffilmark{3}}  \author{Michael M. Dunham\altaffilmark{7}}  \author{Naomi Hirano\altaffilmark{1}} \author{Thomas Henning\altaffilmark{8}}  \author{Michihiro Takami\altaffilmark{1}} \author{Ruobing Dong\altaffilmark{1,9,10}} \author{Jun Hashimoto\altaffilmark{11}} \author{Yasuhiro Hasegawa\altaffilmark{12}} \author{Carlos Carrasco-Gonz\'{a}lez\altaffilmark{3}}

\affil{$^{1}$Academia Sinica Institute of Astronomy and Astrophysics, P.O. Box 23-141, Taipei, 106 Taiwan}\email{baobabyoo@gmail.com}
\affil{$^{2}$European Southern Observatory (ESO), Karl-Schwarzschild-Str. 2, D-85748 Garching, Germany}
\affil{$^{3}$Instituto de Radioastronom{\'i}a y Astrof{\'i}sica, UNAM, A.P. 3-72, Xangari, Morelia, 58089, Mexico}
\affil{$^{4}$Department of Astrophysics, University of Vienna, Tuerkenschanzstrasse 17, 1180, Vienna, Austria}
\affil{$^{5}$Research Institute of Physics, Southern Federal University, Rostov-on-Don, 344090, Russia}
\affil{$^{6}$Konkoly Observatory, Research Centre for Astronomy and Earth Sciences, Hungarian Academy of Sciences, PO Box 67, 1525 Budapest, Hungary}
\affil{$^{7}$Harvard-Smithsonian Center for Astrophysics, 60 Garden St, MS 78, Cambridge, MA 02138}
\affil{$^{8}$Max-Planck-Institut f\"{u}r  Astronomie K\"{o}nigstuhl, 17 D-69117 Heidelberg}
\affil{$^{9}$Hubble Fellow}
\affil{$^{10}$Department of Astronomy, UC Berkeley, 147 Del Mar Ave, Berkeley, CA, 94708}
\affil{$^{11}$National Astronomical Observatory of Japan, 2-21-1 Osawa, Mitaka, Tokyo 181-8588 Japan}
\affil{$^{12}$Jet Propulsion Laboratory, California Institute of Technology, Pasadena, CA 91109, USA}







\begin{abstract}
We report Submillimeter Array (SMA) 1.3 mm high angular resolution observations towards the four EXor type outbursting young stellar 
objects (YSOs) VY Tau, V1118 Ori, V1143 Ori, and NY Ori.
The data mostly show low dust masses $M_{dust}$ in the associated circumstellar disks.
Among the sources, NY Ori possesses a relatively massive disk with $M_{dust} \sim 9 \times 10^{-4}$ $M_{\odot}$.
V1118 Ori has a marginal detection equivalent to $M_{dust} \sim 6 \times 10^{-5}$ $M_{\odot}$.
V1143 Ori has a non-detection also equivalent to $M_{dust} < 6 \times 10^{-5}$ $M_{\odot}$.
For the nearest source VY Tau, we get a surprising non-detection which provides a stringent upper limit 
$M_{dust} < 6 \times 10^{-6}$ $M_{\odot}$. 
We interpret our findings as suggesting that the gas and dust reservoirs that feed the short duration, repetitive 
optical outbursts seen in some EXors may be limited to the small scale, innermost region of their circumstellar disks. 
This hot dust may have escaped our detection limits. Follow-up, more sensitive millimeter observations 
are needed to improve our understanding of the triggering mechanisms of EXor type outbursts.
\end{abstract}



\section{Introduction }
\label{chap_introduction}
One of the most intriguing phenomena in low-mass young stellar objects (YSOs) are the extreme accretion outbursts 
characterized by a sudden brightening in the optical by 1 to 6 magnitudes (Herbig 1977). 
The two most well known types of accretion outburst objects are, namely, the FU Orionis objects (FUors), which have a long burst 
duration of  $\sim$100 years (Hartmann \& Kenyon 1996); and the EXor objects (EXors), which have a short duration 
($\sim$100 days) and are repetitive (Herbig 1989, 2008).
During the active burst stage, the protostellar accretion rate of the classical T Tauri stars (CTTS) can be enhanced 
from the typical values of $\sim$10$^{-9}$ $M_{\odot}~\mathrm{yr}^{-1}$ to up to $\sim$10$^{-4}$ $M_{\odot}$ yr$^{-1}$.
The observational discovery of these outburst sources has provided the most promising solution to the {\it luminosity problem} in the quasi-stationary paradigm of low-mass star formation. 
This problem consists in that the observed mean accretion luminosity of embedded sources is 10 to 100 times smaller (e.g Kenyon 1993; Evans et al. 2009; Dunham \& Vorobyov 2012) than that required for the star to accrete most of its mass in the short ($\sim 10^5$ years) embedded timescale.
This problem can be solved if the star accretes strongly during the accretion outburst episodes.
While earlier optical surveys detected most of the outbursts from the CTTS, recent optical and infrared monitoring observations have 
found that they are widespread from the Class I to the Class III evolutionary stages (Audard et al. 2014).

\begin{table*}\small
\begin{center}
\caption{\footnotesize{Summary of the SMA observations.}}
\label{tab:obs}
\hspace{-1cm}
\begin{tabular}{  p{3.3cm}  p{2.5cm} p{2.5cm} p{2.5cm} p{2.5cm} p{2.5cm}   }\hline\hline
Observing Dates  		&  2013Nov.09  	& 	2013Nov.22  &  2015Jan.22  & 2015Jan.26	&  2015Jan.27 \\
Array Configuration	&  extended  &  extended  &  very extended   & very extended & very extended \\
Number of Antennas &  7  &  7  &  7   &  5    &   6  \\
$uv$ range ($k\lambda$) & 30-170  & 25-175  &  25-390  &  25-350  &  25-350 \\
$\tau_{\mbox{\tiny{225 GHz}}} $  & 0.4 &  0.1-0.4  & 0.07  & $\lesssim$0.1  &  0.07   \\
IFs (GHz) &  4-8 & 4-8  & 4-8, 8-9.5, 10.5-12 & 4-8, 8-9.5, 10.5-12 &  4-8, 8-9.5, 10.5-12 \\
Target sources &	VY\,Tau	& VY\,Tau	&	 XZ\,Tau, VY\,Tau, V1118\,Ori, V1143\,Ori  &  VY\,Tau, V1118\,Ori, V1143\,Ori  &  VY\,Tau, V1118\,Ori, V1143\,Ori  \\
Flux calibrator & Uranus	&	Callisto &	Callisto	& Callisto	& Callisto \\
\hline
Observing Dates  		& 2015Feb.03  &  2015Mar.08  &  2015Apr.02  &  2015Apr.04  &  2015 Apr.08\\
Array Configuration	& extended  & extended & extended  & extended & extended \\
Number of Antennas & 6  &  6 &  6  &  7  & 5 \\
$uv$ range ($k\lambda$) &  25-175  & 25-175  & 30-170 & 30-170  &  20-140 \\
$\tau_{\mbox{\tiny{225 GHz}}} $  &  0.3   & 0.3-0.4  & 0.3-0.4 &  0.1-0.2  &  0.15\\
IFs (GHz) & 4-8 & 4-8 & 4-8 & 4-8 & 4-8 \\
Target sources &  VY\,Tau, V1118\,Ori, V1143\,Ori   &   VY\,Tau, V1118\,Ori, V1143\,Ori   &   NY\,Ori, V1118\,Ori, V1143\,Ori    & NY\,Ori, V1118\,Ori, V1143\,Ori  & NY\,Ori, V1118\,Ori, V1143\,Ori
\\
Flux calibrator &	Callisto	& Ganymede	&	Ganymede	& Callisto	& Titan	\\
\hline
\end{tabular}
\end{center}
\vspace{0.7cm}
\end{table*}
\normalsize{}

\begin{table*}\small
\begin{center}
\caption{\footnotesize{Summary for individual target sources}}
\label{tab:targets}
\hspace{-1.5cm}
\begin{tabular}{  l  p{2.5cm} p{3.1cm} p{2.5cm} p{3.3cm}   }\hline\hline
Source name	&  VY Tau	& NY Ori		& 	V1143 Ori		&	V1118	Ori \\\hline
Evolutionary Class					&	Class III	& Class II	&  Class II			&  Class II		\\
stellar R.A. (J2000)	&	04$^{\mbox{\scriptsize{h}}}$39$^{\mbox{\scriptsize{m}}}$17$^{\mbox{\scriptsize{s}}}$.412	& 	05$^{\mbox{\scriptsize{h}}}$35$^{\mbox{\scriptsize{m}}}$36$^{\mbox{\scriptsize{s}}}$.011 &	05$^{\mbox{\scriptsize{h}}}$38$^{\mbox{\scriptsize{m}}}$03$^{\mbox{\scriptsize{s}}}$.890 & 05$^{\mbox{\scriptsize{h}}}$34$^{\mbox{\scriptsize{m}}}$44$^{\mbox{\scriptsize{s}}}$.745 	\\
stellar Decl. (J2000)	&	+22$^{\circ}$47$'$53$\farcs$40		&	$-$05$^{\circ}$12$'$25$\farcs$31	&	$-$04$^{\circ}$16$'$42$\farcs$81  & $-$05$^{\circ}$33$'$42$\farcs$18	\\ 
Spectral type$^{5}$		&	M0 (with a M2-M4 companion, orbital period $>$350 yr)	&	K	&	M2	&	M2-M3 (with a companion of unclear spectral type, separation$\sim$76 AU)\\
Onset (yr)							&	many (1900-1970, 2013-present) 	&	many	&	many	&	many\\
Outburst duration	 (yr)$^{1}$			&	0.5-2	&	$>$0.3	&	$\sim$1	&	$\sim$1.2\\
Accretion Rate$^{1}$				&	$\cdots$ 	&	$\cdots$	&	$\cdots$ 	&	2.5e-7 (L), 1e-6 (H)\\
Assumed distance (pc)		&	140	&	420	&	420	&	420\\\hline
Synthesized beam & 	$0\farcs59$$\times$$0\farcs40$; 75$^{\circ}$	&	$2\farcs0$$\times$$0\farcs86$; 64$^{\circ}$	&	$0\farcs58$$\times$$0\farcs47$; 65$^{\circ}$  &  $0\farcs61$$\times$$0\farcs47$; 70$^{\circ}$	\\
 ($\theta_{\mbox{\scriptsize{maj}}}$$\times$$\theta_{\mbox{\scriptsize{min}}}$; P.A.)	 & \\
Image RMS (mJy\,beam$^{-1}$)		&	0.55	&	1.7	&	0.62	&	0.60	\\\hline
mm R.A. (J2000)$^{2}$	& 	$\cdots$	&	05$^{\mbox{\scriptsize{h}}}$35$^{\mbox{\scriptsize{m}}}$36$^{\mbox{\scriptsize{s}}}$.013$\pm$$0\farcs05$ & $\cdots$ &	05$^{\mbox{\scriptsize{h}}}$34$^{\mbox{\scriptsize{m}}}$44$^{\mbox{\scriptsize{s}}}$.753$\pm$$0\farcs03$\\
mm Decl. (J2000)$^{2}$	&	$\cdots$	&	$-$05$^{\circ}$12$'$25$\farcs$28$\pm$$0\farcs02$ & $\cdots$	&	$-$05$^{\circ}$33$'$42$\farcs$27$\pm$$0\farcs07$ \\
Image component size$^{3}$  & $\cdots$ & $1\farcs9$$\pm$$0\farcs12$$\times$$0\farcs92$$\pm$$0\farcs03$; 67$^{\circ}$$\pm$1$^{\circ}$.5  & $\cdots$	& $0\farcs69$$\pm$$0\farcs16$$\times$$0\farcs37$$\pm$$0\farcs05$; 17$^{\circ}$$\pm$7$^{\circ}$.6\\
($FWHM{\mbox{\scriptsize{maj}}}$$\times$$FWHM{\mbox{\scriptsize{min}}}$; P.A.) & \\
Peak intensity (mJy\,beam$^{-1}$)	&	$\cdots$	&	28$\pm$1.2 & $\cdots$		&	2.3$\pm$0.4	\\
Peak S/N		&	$\cdots$ & 	16	&  $\cdots$ &	3.8	\\
Integrated 1.3 mm Flux (mJy)	&	3$\sigma$$<$1.7	&	28$\pm$2.2	&	3$\sigma$$<$1.9 &  2.0$\pm$0.7\\\hline
Dust mass$^{4}$ (10$^{-5}$ $M_{\odot}$)	&	3$\sigma$$<$0.58$^{+0.54}_{-0.28}$ 	& 87$^{+93}_{-46}$ & 3$\sigma$$<$5.8$^{+5.5}_{-2.8}$	&	6.2$^{+9.8}_{-4.1}$ \\\hline

\end{tabular}
\end{center}
\footnotesize{$^{1}$Quoted from Audard et al. (2014).}\par \vspace{0.1cm}
\footnotesize{$^{2}$Positions of the detected millimeter sources, derived based on 2-dimensional Gaussian fittings.}\par \vspace{0.1cm}
\footnotesize{$^{3}$Sizes of the detected image components, based on 2-dimensional Gaussian fittings.}\par 
\footnotesize{\hspace{0.1cm}All observed sources are unresolved and therefore cannot be deconvolved. 
These values can be considered upper limits of disk sizes.}\par \vspace{0.1cm}
\footnotesize{$^{4}$Ranges are given based on the assumption of optically thin, $T_{d}$=30 K, and $\beta$=1.0$\pm$0.5.}\par
\footnotesize{\hspace{0.1cm}An upper limit of gas$+$dust mass can be derived assuming a gas to dust mass ratio of 100.}\par \vspace{0.1cm}
\footnotesize{$^{5}$Discussion of spectral types see Cohen \& Kuhi(1979), Hillenbrand (1997), Parsamian et al. (2002), Herbig (2008).}\par
\footnotesize{\hspace{0.1cm}Information about companion see Leinert et al. (1993), Woitas et al.(2001), Reipurth et al. (2007), Dodin et al. (2015).}\par\vspace{0.1cm}
\vspace{0.6cm}
\end{table*}
\normalsize{}

There are well established theories for the outburst triggering mechanisms, including gravitational instability and disk fragmentation
(GI+DF, e.g. Vorobyov 2013; Vorobyov \& Basu 2015, and references therein), the combined effect of GI and magnetorotational instabilities 
(GI+MRI, e.g. Zhu et al. 2009b; Bae et al. 2014), thermal instabilities in the inner disk 
(Lin, Papaloizou \& Faulkner 1985; Bell \& Lin 1994; Hirose 2015), planet-disk interactions (Nayakshin \& Lodato 2012), 
and the encounters of stellar companions (Pflazner et al. 2008).
Most of the existing theories have been initially developed to explain FUor outbursts. 
Recent observations 
have suggested that EXors and FUors may be a similar phenomenon but at extremes of the parameter space (K{\'o}sp{\'a}l et al. 2011b).
All theories but the planet-disk interactions cannot easily explain the short duration and duty cycle of EXors. 
On the other hand, the discovery of a long duration burst from a Class 0 YSO (Safron et al. 2015, Galv\'{a}n-Madrid et al. 2015),  
which is likely earlier than the epoch of planet-formation, disfavors that planet-disk interaction is the triggering mechanism for 
 at least some FUor bursts.
The reported limit  of $<$0.02 $M_{\odot}$ for the disk mass of the FUor object HBC722 might be on the lower end for the 
GI+MRI scenarios to occur (Dunham et al. 2012), although it does not strictly forbid it.

(Sub)millimeter wavelength observations to constrain the properties of circumstellar disks and envelopes in YSOs with 
optical bursts can be a crucial step to address the differences between FUors and EXors, and to clarify their possible triggering 
mechanisms. 
K{\'o}sp{\'a}l (2011b) presented interferometric millimeter observations on FUor objects.
We have observed the 4 EXors: VY Tau, V1118 Ori, V1143 Ori, and NY Ori at 1.3 mm using the Submillimeter 
Array\footnote{The Submillimeter Array is a joint project between the Smithsonian Astrophysical Observatory and the Academia Sinica 
Institute of Astronomy and Astrophysics, and is funded by the Smithsonian Institution and the Academia Sinica.} 
(SMA; Ho, Moran, \& Lo 2004).
Details of our observations and data reduction are outlined in Section \ref{chap_obs}.
The results are presented in Section \ref{chap_result}.
A brief discussion is provided in Section \ref{chap_discussion}.


\section{Observations and Data Reduction} 
\label{chap_obs}
We performed the SMA 1.3 mm observations in the extended and the very extended array configurations.
The observations were carried out using a track-sharing strategy.
A summary is provided in Table \ref{tab:obs}.
The pointing and phase referencing centers are the stellar positions reported for the YSOs (see Table \ref{tab:targets}).
The projected baseline lengths covered by our observations are in the range of 25-390 $k\lambda$, which yielded 
an angular resolution of $\sim$$0\farcs5$, and a maximum detectable angular scale (i.e., a recovered flux $\sim$1/$e$ of the 
original) of $\sim$4$''$, or $\sim$560 AU for VY Tau and $\sim$1680 AU for the Orion targets (Wilner \& Welch 1994).
We attempted a fifth target source (XZ Tau) on 2015 January 22, and found that we cannot achieve a good image quality 
due to the strong sidelobe response to the adjacent bright millimeter source HL Tau.

All observations used the Application-Specific Integrated Circuit (ASIC) correlator, which provided the 4-8 GHz intermediate 
frequency (IF) coverages in the upper and lower sidebands, with 48 spectral windows in each sideband. 
The observations tracked the rest frequency of 230.538 GHz at spectral window 22 in the upper sideband.
For our observations on 2015 January 22, 26, and 27, the newly commissioned SMA Wideband Astronomical ROACH2 Machine (SWARM) 
correlator provided two additional spectral windows, which covered the 8-9.5 GHz and 10.5-12 GHz IFs, respectively.
However, we omit using data in the 10.5-12 GHz IF because of poor response.

The application of $T_{\mbox{\scriptsize{sys}}}$ information, and the absolute flux, passband, and gain calibrations were 
carried out using the MIR IDL software package (Qi 2003).
After calibration, the zeroth order fitting of continuum levels, and the joint (Briggs robust=2) weighted imaging of all 
continuum data were performed using the Miriad software package (Sault et al. 1995). 
We do not use the data taken on 2015 April 04, since the large and rapid 
atmospheric phase variations hampered calibration. 
The achieved rms noise levels for the sources can be found in Table \ref{tab:targets}.
A typical absolute flux calibration accuracy is $\sim$15\% for SMA observations.


\section{Results}
\label{chap_result}
\begin{figure*}
\hspace{0.2cm}
\hspace{0.15cm}
\begin{tabular}{p{7.5cm} p{7.5cm} }
\includegraphics[width=8.25cm]{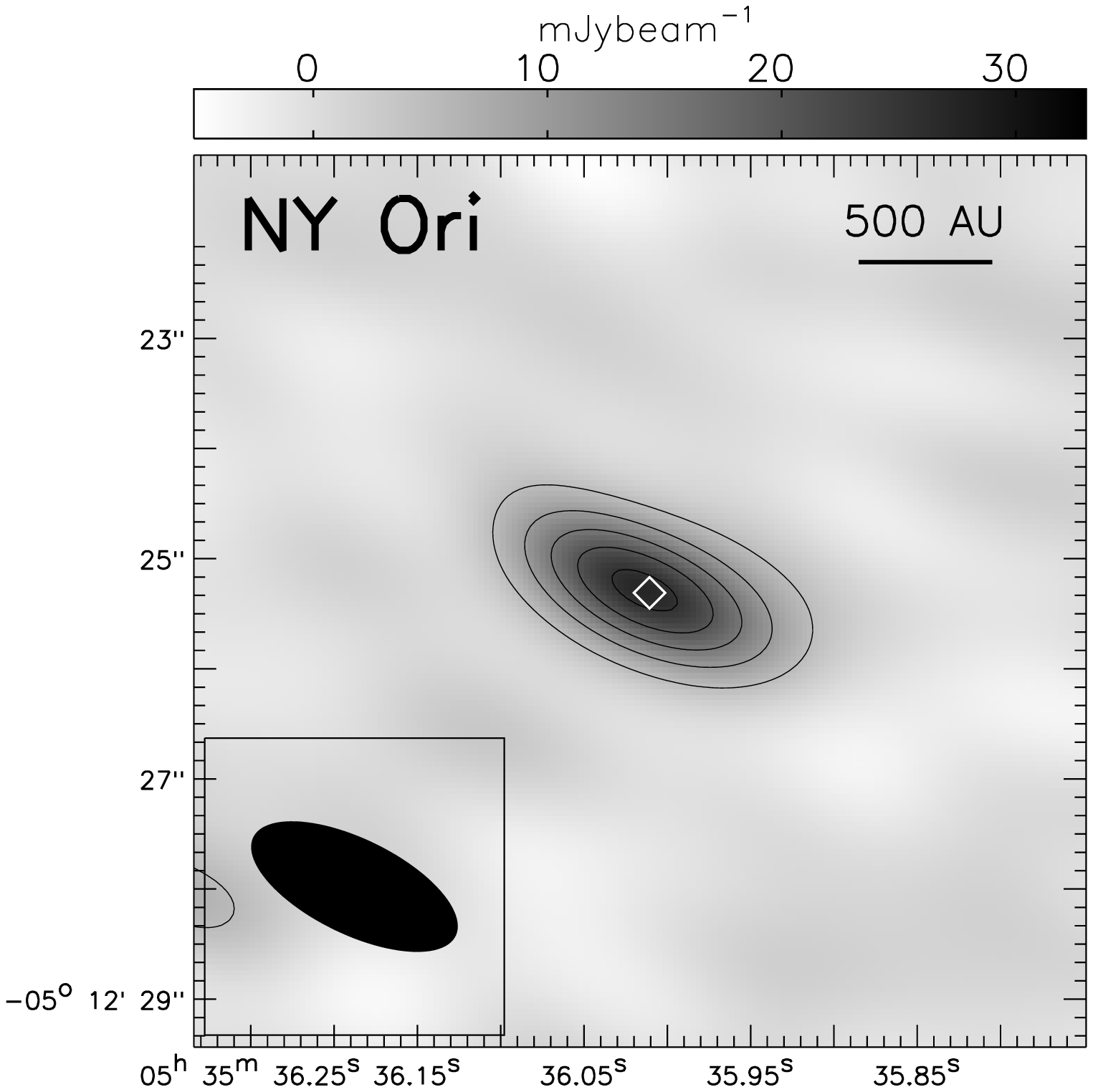} & \includegraphics[width=8.25cm]{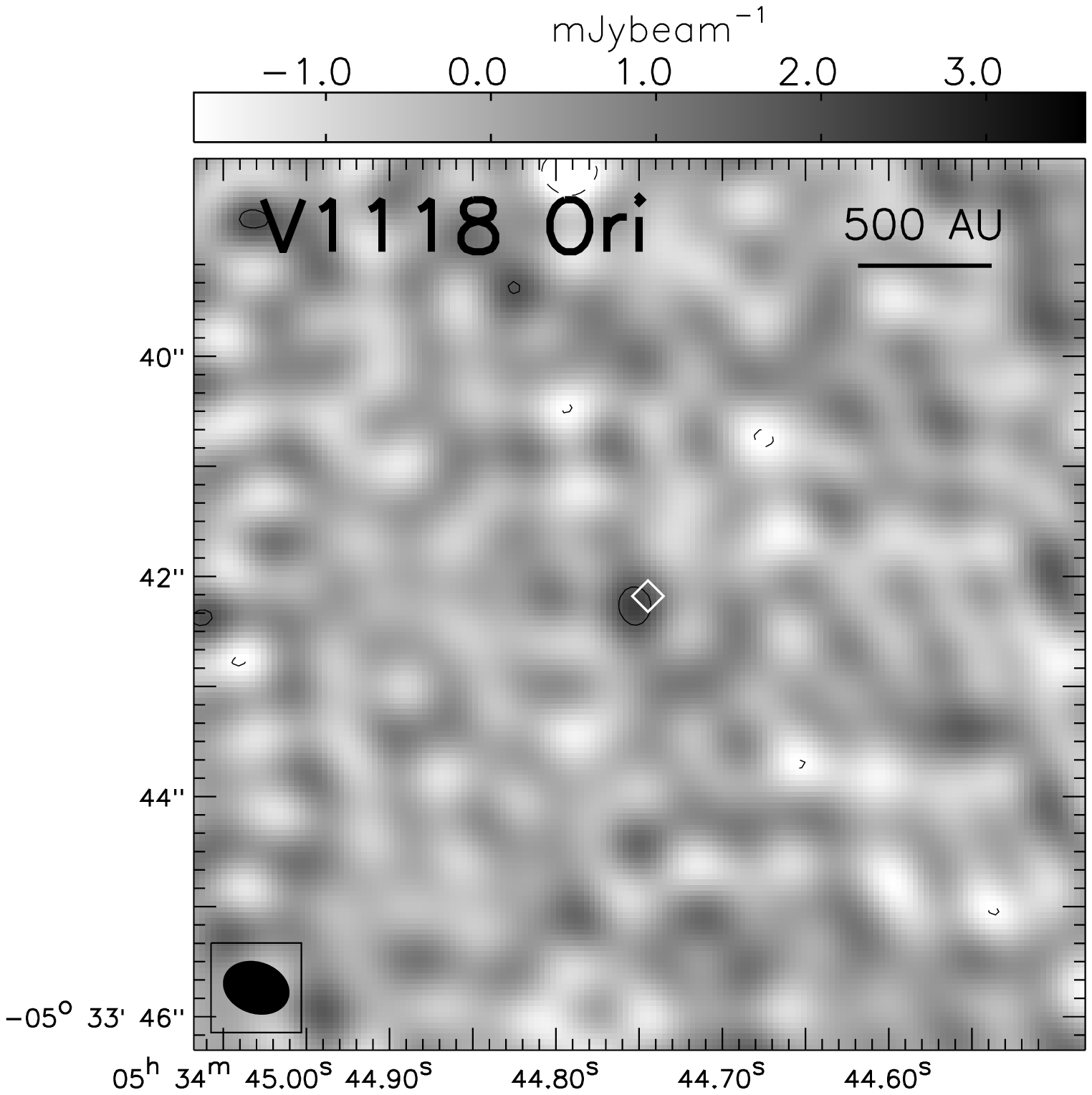} \\
\end{tabular}

\vspace{-1cm}
\hspace{0.3cm}
\begin{tabular}{p{7.6cm} p{7.5cm} }
\includegraphics[width=8.9cm]{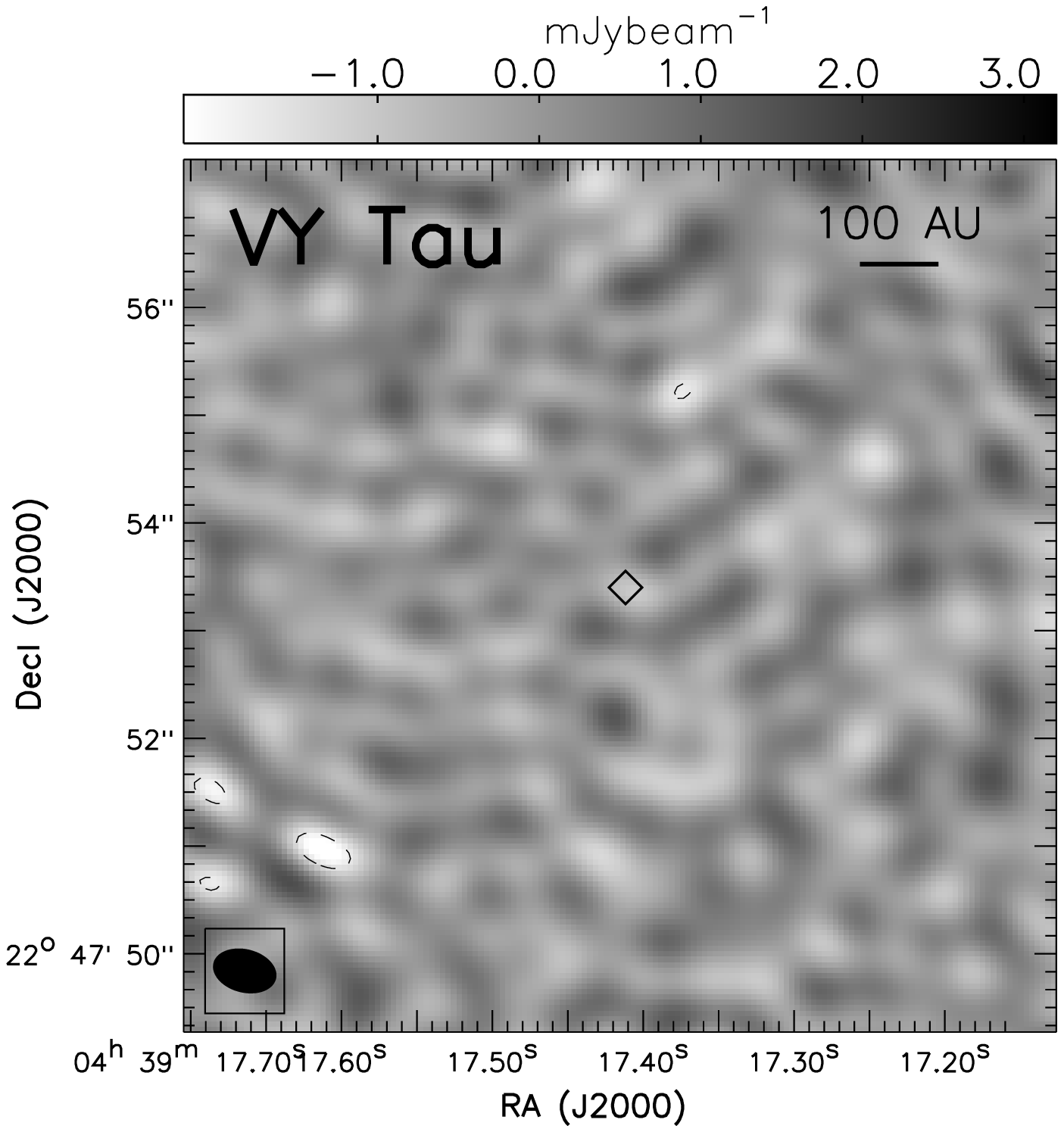} & \includegraphics[width=8.25cm]{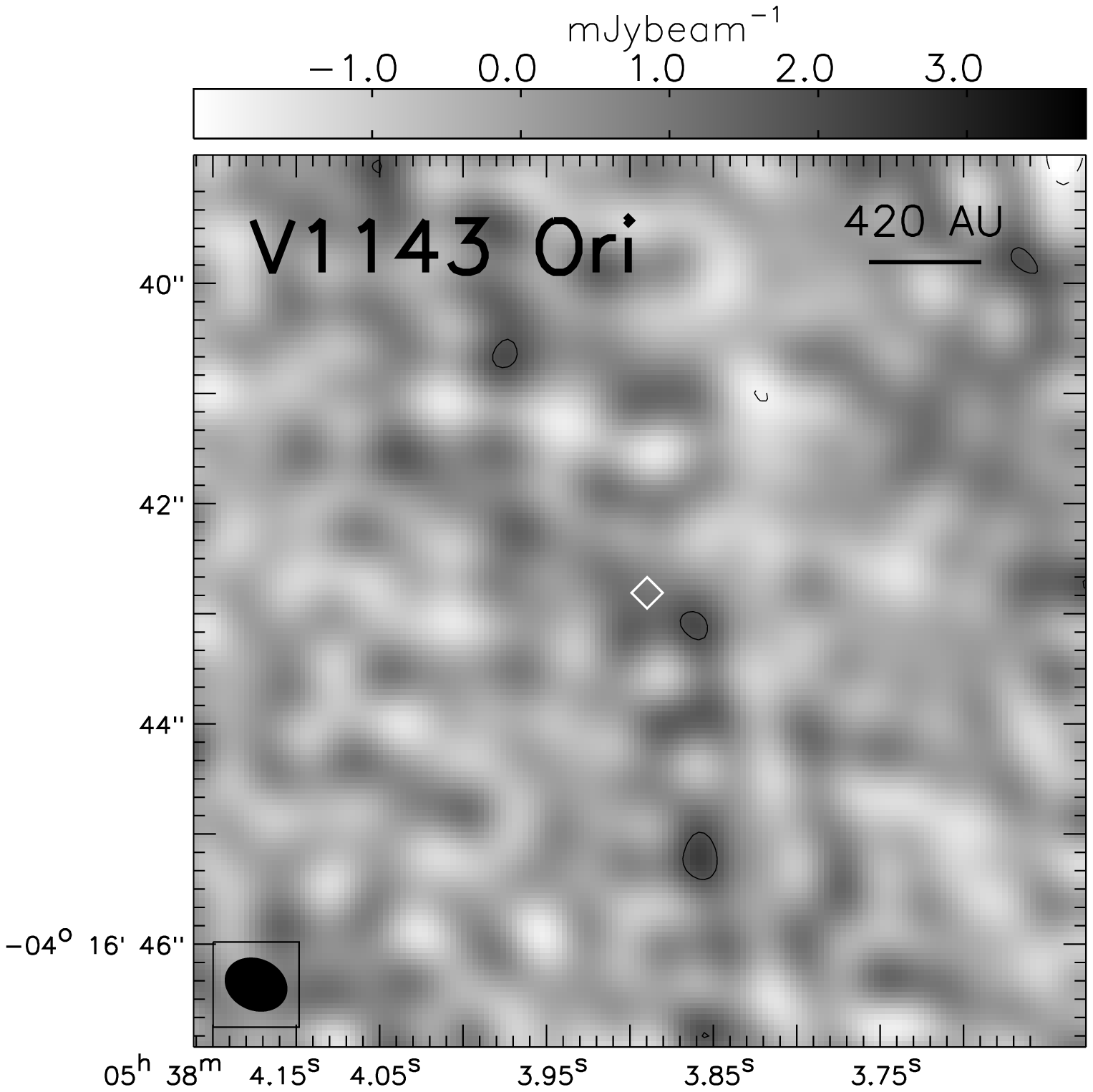} \\
\end{tabular}
\caption{\small{SMA 1.3 mm continuum images of the EXor-type young eruptive stars NY Ori, V1118 Ori, VY Tau, and V1143 Ori (contours and grayscale). 
Diamonds represent the optical stellar positions.
The synthesized beam of each of these images is shown in the bottom left. 
Contours are -3$\sigma$ (dashed), and 3$\sigma$$\times$[1, 2, 3, 4, 5].
The 1$\sigma$ levels can be found in Table \ref{tab:targets}.
}}
\label{fig:images}
\vspace{0.5cm}
\end{figure*}

Figure \ref{fig:images} shows the SMA 1.3 mm continuum images of NY Ori, V1118 Ori, VY Tau, and V1143 Ori.
We significantly detected the dust emission toward NY Ori.
We report a tentative, 3.8$\sigma$ detection for V1118 Ori.
Although some features at 2 to 3$\sigma$ are seen close to V1143 Ori, they are offset from the stellar position. 
Also, some of these faint features look like sidelobe responses to nearby sources, or could be 
related to unidentified, baseline based visibility amplitude errors. Therefore, we assign a non-detection 
(3$\sigma$$<$1.9 mJy) for V1143 Ori. VY Tau is clearly non-detected in our 1.3 mm observations.

The selected YSOs are in the Class II/III evolutionary stages (Audard et al. 2014, and references therein). 
If the circumstellar disks of EXors are not far bigger than those of typical Class II/III YSOs, according to previous 
(sub)millimeter interferometric surveys we do not expect our targets to possess dust disks larger than $\sim 500$ AU  
(e.g., Dutrey et al. 1996, Andrews 2007a, 2009, Carpenter et al. 2014, Williams \& Best 2014). 
Given the $\sim$500-1000 AU maximum recoverable angular scale of our observations (Section \ref{chap_obs}), we do not think 
our non-detections can be explained as due to missing flux from large disks, filtered out by the interferometer. 
If the non-detections were due to missing short-spacing data, the images would present extended stripes or ripples, 
which are not found in our images. 
We attribute the non-detections in VY Tau and V1143 Ori to a low content of cold dust (more in Section \ref{chap_discussion}).
Neither NY Ori nor V1118 Ori is spatially resolved by our observations.

There is no known companion to NY Ori.
V1118 Ori has a close companion $0\farcs18$ away (Reipurth et al. 2007).
Given our angular resolution, we cannot distinguish whether or not the detected millimeter emission has a significant contribution from the 
companion. 
Therefore, the derived dust mass for an accretion disk in the V1118 Ori EXor should be regarded as an upper limit.  

\section{Discussion}
\label{chap_discussion}
For YSOs at the evolutionary stages of our selected sources, free-free emission from ionized jets and photoevaporation winds, 
and gyrosynchrotron emission from the protostars are dim and well below our detection limit at 1.3 mm,  
although gyrosynchrotron emission may have temporal outbursts (e.g., K{\'o}sp{\'a}l et al. 2011a; Liu et al. 2014, Galv{\'a}n-Madrid et al. 2014).
It is safe to assume that the dominant contribution to the 1.3 mm emission is by thermal dust emission from the circumstellar disks.
If gyrosynchrotron emission contributes significantly, the millimeter flux may vary significantly on timescales of days (e.g., K{\'o}sp{\'a}l et al. 2011a).
We did not observe evidence of millimeter flux variability, although this can be due to limited sensitivity. 
In the case that gyrosynchrotron emission does contribute, our observations will provide upper limits for dust masses.

The dust mass $M_{dust}$ inferred from the 1.3 mm observations can be estimated based on the optically thin formula
\[
M_{dust} = \frac{2\lambda^{3}a\rho D^{2}}{3hcQ(\lambda)J(\lambda,T_{d})}S(\lambda),
\]
where $a$ is the mean grain radius, $\rho$ is the mean density of dust grains, $D$ is the distance of the target, 
$Q$($\lambda$)$\propto$$\lambda^{-\beta}$ is the grain emissivity, $T_{d}$ is the dust temperature, $S(\lambda)$ is the flux 
at the given wavelength, $J(\lambda, T_{d})=1/[\mbox{exp}(hc/\lambda k_{B}T_{d})-1]$ (Hildebrand 1983; Lis et al. 1998).
$c$, $h$, and $k_{B}$ are the light speed, Planck constant, and Boltzmann constant, respectively. 
The derived dust mass, adopting the typical assumption of $a$=0.1 $\mu$m, $\rho$=3 g\,cm$^{-3}$, $\beta$=1.0$\pm$0.5, $Q$($\lambda$=350 $\mu$m)=1$\times$10$^{-4}$ and a dust temperature $T_{d}$$\sim$30 K, are listed in Table \ref{tab:targets}.
Our assumptions effectively extrapolate dust opacity from shorter wavelength bands, and incorporate the effects of grain growth into the opacity index $\beta$.
Our expression is equivalent to an assumed dust opacity of 0.67$^{+0.62}_{-0.32}$ cm$^{2}$g$^{-1}$ at 230 GHz.
The adopted range of dust opacity well incorporates the expected values from theories, for $\sim$1-100 mm maximum grain radius (i.e. the assumed lowest opacity 0.35 corresponds to the case of $\sim$10 cm maximum grain radius. c.f. Draine 2006).
We refer to Isella et al. (2009) and Guilloteau et al. (2011) for a discussion of dust temperatures at the corresponding 
spatial scales. 
We refer to Guilloteau et al. (2011) and  Ricci et al. (2010, 2012) for the commonly observed 
range of $\beta$ in circumstellar disks. 
The derived dust masses, including upper limits, are provided in Table \ref{tab:targets}.

The effect of different grain sizes can be estimated using the relation (Beckwith et al. 1991):

$${{Q(\lambda)} \over {a}} = {{4 \rho \kappa(\lambda)} \over {3}},$$

\noindent where $\kappa(\lambda)$ is the mass absorption coefficient as a  function of grain radius. 
This last parameter has been calculated by D'Alessio et al. (2001; their Figure 3) assuming  a size distribution of dust grains taken to be a power law, $n(a) \propto a^{-p}$ , with $p$ in the range of 2.5 to 3.5, and a  maximum grain radius $a_{max}$.  
Adopting $p$ = 3.5, we find that the ratio $Q(\lambda)/a$ for observations made at $\lambda$ = 1.3 mm shows three regimes as a function of $a_{max}$. 
For $a_{max} \leq 10~\mu m$ the ratio remains constant and the derived dust masses are not affected. 
For $10~\mu m \leq a_{max} \leq 10~cm$, $Q(\lambda)/a$ first rises with increasing $a_{max}$ to reach at $a_{max}$ = 0.1 cm a value about 8 times larger than that in the $a_{max} \leq 10~\mu m$ regime. 
Then the ratio decreases again to reach a value similar to that of the $a_{max} \leq 10~\mu m$ regime at $a_{max}$ = 10 cm. 
Then, in this size range, the derived dust masses can actually be smaller than those derived by us in the small size limit. 
For the specific case of $a_{max}$ = 10 cm, the derived dust mass is approximately equal to that derived in the small size limit and our mass determinations or upper limits remain valid. 
Finally, for $a_{max} \gg 10~cm$ the $Q(\lambda)/a$ ratio starts to decrease, implying that the derived values and upper limits of the dust mass derived by us will start growing. 
Then, for the case of very large grain sizes the amount of dust mass could be substantially larger that estimated here.




The derived values and upper limits for $M_{dust}$ in our sample span two orders of magnitude. 
With the exception of NY Ori, it seems that the disk (or disk plus inner envelope) masses of EXors are systematically lower than those of FUor objects 
(K{\'o}sp{\'a}l 2011; Dunham \& Liu in prep.), although our small sample may be biased. 
EXors in general seem to be more evolved than FUors, which also agrees with EXors' seemingly lower disk masses, although the difference in evolutionary phases between the two eruptive classes has recently started to wash out (Audard et al. 2014).
Despite the similar masses and spectral types of the host stars (Table \ref{tab:targets}; also see Hartigan \& Kenyon 2003), 
from our millimeter observations we cannot yet find a feature in common among the circumstellar disks of the observed EXors.
In particular, the non-detection of the nearest source VY Tau places very stringent limits to the dust and gas masses.
Considering an upper limit of the gas-to-dust mass ratio of 100 implies that the gas mass of the disk in VY Tau 
is $\ll$10$^{-3}$ $M_{\odot}$, which is 5 times lower than the mean gas mass of the T Tauri disks in previous surveys 
(Andrews et al. 2005; Williams \& Cieza 2011).
The actual gas-to-dust mass ratio of VY Tau is likely much lower than 100 given its late evolutionary stage.
We find the small gas/dust mass in VY Tau to be counterintuitive, because gravitational disk instabilities are more likely 
to occur in massive disks.
NY Ori has a sufficiently massive disk, such that GI+DF or GI+MRI are feasible triggering mechanisms for the accretion bursts.
However, the short burst durations are not yet explained by current models.
The mass estimates for the other three sources also appear to be too low to suggest GI-related mechanisms.

Recent astrometric and photometric monitoring observations of VY Tau (and its companion) suggest that its outbursts are likely not 
triggered by encounters with the companion, which has a $>$350 yr orbital period (Dodin et al. 2015).
NY Ori and V1143 Ori do not have a know companion, so a stellar encounter as their outburst triggering mechanism is disfavored. 
The projected separation between V1118 Ori and its companion is $\sim$76 AU (Table \ref{tab:targets}), thus the 
orbital period of V1118 Ori and its companion should be $> 100$ yr.
It seems that the outbursts of V1118 Ori, which have a short duration and duty cycle, are not triggered by the encounters with its 
companion.

The triggering mechanisms for the short duration, repetitive accretion bursts may be related to phenomena on the scales of 
the inner accretion disk (e.g., D'Angelo \& Spruit 2012, K{\'o}sp{\'a}l et al. 2014, Hirose 2015). 
Whether the specific mechanism is interaction with planetary companions or not, and whether the outburst triggering mechanisms 
are also linked with the disk mass dispersal, remain as intriguing problems.




\acknowledgments
HBL thanks the support from ASIAA and the SMA staff.
HBL thanks T. Muto and S. Hirose for useful discussion.
This research was done with the support of program UNAM-DGAPA-PAPIIT IA101715. 
This work was supported by the Momentum grant of the MTA CSFK Lend\"ulet Disk Research Group.
EIV acknowledges the support from the Russian Ministry of Education and Science Grant 3.961.2014/K and RFBR grant 14-02-00719.
MMD acknowledges support from the Submillimeter Array through an SMA Postdoctoral Fellowship.
Y.H. is supported by JPL/Caltech.

{\it Facilities:} \facility{SMA} 

\end{document}